\documentstyle[]{laa}
 
\begin{document}
 \input epsf
 \newcommand{\dix}{\,{ \,10} }
 \newcommand{\mic}{\,{ \mu m} }
 \newcommand{\pac}{\,{ pc} }
 \newcommand{\Wmsr}{\,{ Wm^{-2}sr^{-1}} }
 \newcommand{\Wmsrmic}{\,{ Wm^{-2}sr^{-1}\mic^{-1}} }
  \newcommand{\Wmsrhz}{\,{ Wm^{-2}sr^{-1}Hz^{-1}} }
 \newcommand{\Wcmsr}{\,{ Wcm^{-2}sr^{-1}} }
 \newcommand{\Wcm}{\,{ Wcm^{-2}} }
  \newcommand{\cm}{\,{ cm^{-2}} }
  \newcommand{\cmmun}{\,{ cm^{-1}} }
 \newcommand{\Wm}{\,{ Wm^{-2}} }
 \newcommand{\WH}{\,{ W/H_{atom}} }
 \newcommand{\Wcmmic}{\,{ Wcm^{-2}\mic^{-1}} }
 \newcommand{\Wmmic}{\,{ W/m^2/\mic} }
 \newcommand{\MJysr}{\,{ MJysr^{-1}} } 
 \newcommand{\Wcmmicsr}{\,{ W/cm^2/\mic /sr} }
 \newcommand{\Dl}{\,{ \Delta \lamb} }
 \newcommand{\lamb}{\,{ \lambda} }
\newcommand{\Inu}{\,{ I_{\nu}}}
 \newcommand{\nuInu}{\,{ \nu I_{\nu}}}
 \newcommand{\nHdeux}{\,{n_{H_2}} }
\newcommand{\nHtwo}{\,{n_{H_2}} }
\newcommand{\NHdeux}{\,{N_{H_2}} }
\newcommand{\nHII}{\,{n_{HII}} }
\newcommand{\nH}{\,{n_{H}} }
\newcommand{\NH}{\,{N_{H}} }
\newcommand{\cmcube}{\,{ cm^{-3}} }
\newcommand{\cmdeux}{\,{ cm^{-2}} }
\newcommand{\cmdeuxkms}{\,{ cm^{-2}(km/s)^{-1}} }
\newcommand{\cmun}{\,{ cm^{-1}} }
\newcommand{\cmcubes}{\,{ cm^{-3}s^{-1}} }
\newcommand{\kcmcubes}{\,{ cm^{3}s^{-1}} }
\newcommand{\tUV}{\,{\tau_{\mbox{UV}}} }
\newcommand{\tV}{\,{\tau_{\mbox{V}}} }
\newcommand{\Lsol}{\,{ L_{\sun}} }
\newcommand{\IO}{\,{ I_0} }
\newcommand{\IL}{\,{ I_L} }
\newcommand{\epstrois}{\, {\varepsilon_{3.3}} }
\newcommand{\Hbeta}{\,{ H_{\beta}} }
\newcommand{\Halpha}{\,{ H_{\alpha}} }
\newcommand{\Bralpha}{\,{ Br_{\alpha}} }
\newcommand{\pccmsix}{\,{ pc \, cm^{-6}} }
\newcommand{\Htwo}{\,{ H_2} }
\newcommand{\CeighteenO}{\,{C^{18}O} }
\newcommand{\Av}{\,{ A_V} }
\newcommand{\Cplus}{\,{ C^+} }
\newcommand{\Etrois}{\, {E_{3.3}} }
\newcommand{\FC}{\, {F_{\lambda}(C)} }
\newcommand{\FK}{\, {F_{\lambda}(K)} }
\newcommand{\FL}{\, {F_{\lambda}(L)} }
\newcommand{\FPAH}{\, {F_{\lambda}(PAH)} }
\newcommand{\hnu}{\, { h\nu}}
\newcommand{\douzeCO}{\, { ^{12}CO}}
\newcommand{\treizeCO}{\, { ^{13}CO}}
\newcommand{\Hplus}{\, { H^+}}
\newcommand{\HdeuxO}{\,{ H_{2}0}}
\newcommand{\NHtrois}{\,{ NH_{3}}}
\newcommand{\orthoNHtrois}{\,{ ortho-NH_{3}}}
\newcommand{\paraNHtrois}{\,{ para-NH_{3}}}
\newcommand{\HtwoeighteenO}{\,{ H_{2}^{18}O}}
\newcommand{\HtwoO}{\,{ H_{2}O}}
\newcommand{\orthoHtwoO}{\,{ortho-H_{2}O}}
\newcommand{\paraHtwoO}{\,{para-H_{2}O}}
\newcommand{\Odeux}{\,{ O_{2}}}
\newcommand{\Otwo}{\,{ O_{2}}}
\newcommand{\Fd}{\,{ F_d}}
\newcommand{\betal}{\,{ \beta_{l}}}
\newcommand{\betad}{\,{ \beta_{d}}}
\newcommand{\gu}{\,{ g_{u}}}
\newcommand{\gl}{\,{ g_{l}}}
\newcommand{\Eu}{\,{ E_{u}}}
\newcommand{\El}{\,{ E_{l}}}
\newcommand{\Aul}{\,{ A_{ul}}}
\newcommand{\nup}{\,{ n_{u}}}
\newcommand{\nl}{\,{ n_{l}}}
\newcommand{\smun}{\,{ s^{-1}}}

    \thesaurus{Sect. 8         
              ( 09.04.1; 
              09.13.2; 
               09.03.1;  
               09.18.1;   
               13.09.4;     
               11.19.3       
                 )
             }
   \title{Excitation of interstellar molecules by near infrared PAH photons}
 
   \subtitle{}

   \author{M. Giard	\inst{1}
          , J.L. Puget	\inst{2}
          , E. Cr\'et\'e \inst{1}
	  , F. Scoupe 	\inst{1}
          }
 
   \offprints{M. Giard, giard@cesr.cnes.fr}
 
   \institute{
Centre d'Etude Spatiale des Rayonnements, 9 avenue du Colonel Roche, BP
4346, F-31029 Toulouse Cedex, France.
\and Institut d'Astrophysique Spatiale, Campus d'Orsay, B\^at. 121, 
F-91405 Orsay cedex, France.}
 
   \date{Received               ; accepted              }
 
   \maketitle
 
   \begin{abstract}

We have developed an excitation model for small molecules
including radiative pumping by 
dust photons from near infrared to submillimeter wavelengths. This model 
applies to molecules 
within bright photodissociation regions in galactic  
star-forming regions
and starburst galaxies. 
In such environments, the near infrared photons
emitted by polycyclic aromatic hydrocarbon molecules (PAHs) 
or small carbon grains, 
$5 \mic < \lambda < 20 \mic$, are able to penetrate molecular regions
and pump the molecules
in an excited vibration state. The far infrared and submillimeter 
transitions 
involved in the de-excitation cascade are strongly affected by this process. 
Their intensities can be enhanced by several orders of magnitude. 
We have applied this model to $\HtwoO$ and
$\NHtrois$, 
ortho and para species. 
 The behaviour of the FIR rotation lines
with respect to the gas density and the molecule column
density is strongly modified compared to pure collisional models. 
At moderate densities, $\nHdeux < 10^5 \cmcube$, 
radiative pumping dominates the excitation of the rotation ladder, and 
the FIR lines are good probes of the molecule column density.  
The near infrared absorption and emission lines predicted 
to appear between 6 and 7
$\mic$ for $\HtwoO$ and from 10 to 12 $\mic$ for $\NHtrois$ can also
be used for this purpose. 
Concerning $\orthoHtwoO$, a quasi resonant 
pumping of the $\nu_2$ $6.18 \mic$ transition by photons of the 6.2 $\mic$ PAH 
band occurs. A strong de-excitation emission line
is expected at 6.64 $\mic$.


   \end{abstract}
 
%
\section{Introduction}

Modelling the formation of molecular rotation lines
in interstellar clouds is a crucial task both for theoretical
and observational astrophysics. On the theoretical side, early studies by
Goldreich and Kwan (\cite{Goldreich74}) and Goldsmith and Langer 
(\cite{Goldsmith78}) have shown that the rotation lines from
polar and/or abundant molecules like $CO$, $\Otwo$, $HD$, $\HtwoO$ and other
hydrides 
should be the major cooling agents of the molecular gas. Together with 
the heating mechanisms (e.g. cosmis rays, $\Htwo$ formation, 
gravitational contraction and grain heating), they control the thermal 
balance of dense clouds, and thus, the conditions that will allow the
contraction
toward star formation. On the observational side, millimeter and
submillimeter molecular lines are used to probe several physical parameters of 
molecular clouds, such as the gas temperature, the $\Htwo$ density, the
abundance 
of molecules, and the gas dynamics. Densities and temperatures in
molecular 
clouds are such that the collision rate with $\Htwo$ molecules generally does
not
allow to reach the  thermodynamic equilibrium of the rotation ladder
(see Goldreich and Kwan \cite{Goldreich74}). Instead, the coupling with the
radiation field is important, both via absorption and spontaneous emission. 
Takahashi et al. (\cite{Takahashi83} and \cite{Takahashi85}), have shown  
how far infrared dust photons, $\lambda = 30$ to $100 \mic$ are 
able to pump the rotation ladder of the $\HtwoO$
molecule within warm molecular cores.
Carroll and Goldsmith (\cite{Carroll81}) have investigated 
the pumping of molecules via their near infrared (NIR) vibration transitions. 
At that time it was assumed that a significant fraction of the dust emission
could be in the NIR range, only in the immediate vicinity
of hot stars. However, the effect of the NIR 
radiation attributed to large polycyclic aromatic hydrocarbon 
molecules (PAHs) and very small grains, which 
makes up to 30 \% of the total dust emission  (Puget et al. \cite{Puget85}), 
has never been taken into account. This radiation is characterised by a strong
continuum and broad emission features, located at 6.2 and 7.7 $\mic$ for the
strongest ones, wavelengths which co\"{\i}ncide with the vibration transitions 
of
some
molecules. 
For instance the $v2$ transition of $\HtwoO$ and the $v4$ transition of
$\NHtrois$ 
occur at the maximum of the 6.2 $\mic$ PAH feature, and the 7.8 $\mic$ vibration of CS
falls within the wavelength range of the 
7.7 $\mic$ feature. 
There is now clear observational evidence
that the NIR emission bands are not 
only present in peculiar reflexion nebulae, but are a 
general component of the dust radiation:
 detection of the 3.3 and 6.2 $\mic$ bands  in our galaxy
by Giard et al. \cite{Giard94a} and Ristorcelli et al. \cite{Ristorcelli94},
 and, more recently, the measurements performed by the ISO satellite
 (Boulanger et al. \cite{Boulanger96},
 Mattila et al. \cite{Mattila96}, and Verstraete et al. \cite{Verstraete96}). 
The aim of this paper is to revisit the problem of the coupling between
molecules and the dust radiation field, in the framework of a model which 
includes
the PAH emission bands. The mechanism, and the rates involved in some
representative astrophysical environments, are presented in 
Sect. \ref{NIRpump}. A detailed model for the molecular excitation
in star-forming photodissociation regions (PDR) is develloped in Sect.
\ref{Model}.
The results for two molecules, $\HtwoO$ and $\NHtrois$, 
are presented in Sect. \ref{Results}.

\section{\label{NIRpump}  Near infrared pumping of rotation transitions}

Near infrared (NIR) PAH photons, continuum or bands, can excite the vibrations of
a molecule.
This absorption does also imply some rotational excitation ($\Delta$J $\neq$ 0).
After spontaneous de-excitation of the vibration mode, which occurs
almost immediately ($Aij \simeq 10 \smun$), the molecule can be left in an 
excited rotation state. Subsequent excitation from this state to an
upper one can follow via the same mechanism.
We have listed in Table \ref{table:molecules} some molecules of astrophysical
interest which may be affected by this process, because
one of the vibration transitions happen to be at a wavelength
larger than 6 $\mic$ ($< 1670 \cmun$), where the PAH emission is strong.

\begin{table}
\caption[]{\label{table:molecules}
Some molecules of astrophysical interest which happen to have 
one or several vibration transitions that can be pumped by PAH photons. 
Frequencies and Einstein coefficients are from the compilation
of Encrenaz et al. \cite{Encrenaz92}.  If not available the
Einstein coefficient has been approximated to $10 s^{-1}$. 
Column 4: indicative abundance relative to hydrogen.
When available, we have used the value from the dust/gas chemical model of 
Shalabiea and Greenberg \cite{Shalabiea94} at a time evolution of 
$10^6$ years. The abundance of $H_{3}O^{+}$ is from Phillips et al. 
\cite{Phillips92}.
Column 5: optical depth in the absorbing vibration transition for
a typical PDR column density: $2\NHdeux = 10^{21}
\cmdeuxkms$.
}

\begin{flushleft}
\begin{tabular}{lllll}
\hline
 Molecule & $\nu_{vib}$ & $A_{vib}$ & Abundance & $\tau$ \\
  &$\cmun$ & $s^{-1} $ &  &  \\
\hline
$\HtwoO$ & $\nu_2$ 1620 & 20 & $10^{-5}$ & 19\\
$\NHtrois$ & $\nu_2$ 950 & 16 & $10^{-6}$ & 7\\
	  & $\nu_4$ 1627 & 5 &  & 0.5\\
$CO_2$  & $\nu_2$ 670&  1.4 & $10^{-6}$ & 2\\
$C_3H_2$ (cyclic) &$\nu_3$ 1277& 10& $2~10^{-8}$& 0.04\\
$CH_4$ & $\nu_4$ 1306 & 2.6 & $10^{-6}$&0.5\\ 
$HCN$ & $\nu_2$ 712 & 1.7 & $10^{-6}$ & 2\\
$HC_3N$ & $\nu_5$ 663 & 4.7 & $10^{-9}$& 0.006\\
$H_3O^+$ & $\nu_4$ 1626& 10 & $3~10^{-9}$  & 0.003\\
$CH_3OH$ & $\nu_4 ... \nu_8, \nu_{10}, \nu_{11}$&  10 & $3~10^{-7}$ & 1.2\\
         & 1033 ... 1478 &  & &\\
$CS$ & 1285 & 10 & $3~10^{-6}$ & 5.5\\
\hline
\end{tabular}
\end{flushleft}
\end{table}

In order to get a first quantitative estimate of the competition 
between  the proposed mechanism (radiative NIR pumping) and collisional excitation,
we have to figure out what are the rates of each elementary process.
Concerning the effect of collisions with $\Htwo$, the deexcitation 
rate is approximately (see Goldreich and Kwan \cite{Goldreich74})  :

\begin{equation}
\frac{k_{col. deexc.}^{-1}}{\smun} \simeq 7 \dix^{-7}  (\frac{T}{100K})^{0.5}
\frac{\nHdeux}{\dix^4 \cmcube} 
\label{eq:k_col_deexc}
\end{equation}

where T is the gas temperature, and $\nHdeux$ the gas density. The excitation
rate
follows from the detailed balance:

\begin{equation}
\frac{k_{col. exc.}^{-1}}{\smun} \simeq  k_{col. deexc.}^{-1} \frac{\gu}{\gl}
exp(-(\Eu-\El)/kT) 
\label{eq:k_col_exc}
\end{equation}

where $\gu$, $\Eu$ (respectively $\gl$, $\El$) are the statistical weight and
the energy
of the upper (respectively lower) level.

The radiative excitation rate is given by the product of the 
Einstein coefficient by the spectral radiation density:

\begin{equation}
\frac{k_{rad}^{-1}}{\smun} \simeq  6.8 \dix^{-7} \frac{\Aul}{s^{-1}} \frac{<\Inu
>}{MJy/sr} (\frac{\nu}{\dix^{12} Hz})^{-3} 
\label{eq:k_rad}
\end{equation}

where $\Aul$ is the spontaneous emission rate, $<\Inu >$ is the average 
spectral density of the radiation field in the line
(average over frequency and solid angle). 
The spectral density at the maximum of the 6.2 and 7.7 $\mic$ PAH features 
is a few $\dix^4$ MJy/sr 
within typical photodissociation regions (PDRs) associated to star forming
complexes 
(e.g. M17, see Giard et al. \cite{Giard92}, and \cite{Giard94b},) or starburst 
galaxies. This implies
that
in such regions and their vicinity,  the radiative
excitation rate to the vibration excited state,  $\simeq 3 \dix^{-6} \smun$, 
can be higher than the collision
excitation rate of the rotation ladder, $\simeq 7 \dix^{-7} \smun$.  
Table  \ref{firstestimates} compares the radiative and collisional
rates obtained in the
case of  the excitation of the $2_{21}$ rotation level of $\orthoHtwoO$, 111
$\cmmun$ above the ground state. Fig. \ref{fig:h2olevels}a shows in detail how
this level
can be excited via radiative excitation at 6.18 $\mic$ 
of the $v2$ $1_{10}$ state, and spontaneous
emission, $A_{ij} \simeq 10 \smun$.
We have considered different astrophysical environnements 
where radiation can play a significant role: a typical HII/$\Htwo$ galactic PDR 
in a region of massive star formation 
(see Tielens and Hollenbach \cite{Tielens85}); a molecular cloud
nearby to this PDR (e.g. M17 SW);  the central regions
of a starbust galaxy (e.g. M82); and cold molecular gas within the galactic
molecular ring. In the last case, the low value of 
the radiation field is balanced by
the poor collisional efficiency at low temperatures, T = 50 K. 
The collision rates are from Green et
al. \cite{Green93}. The figures in this table are only given to illustrate the 
fact that in such astrophysical environments radiative excitation by NIR PAH
photons
efficiently competes with collisional excitation. A correct treatment has to
take into account
the escape probability of the photons from the cloud. This will be done in the
next
section.  

\begin{table}
\caption[]{\label{firstestimates}
Excitation of the $2_{21}$ rotation level of $\orthoHtwoO$, 111 $\cmun$ above
the 
ground level, for different astrophysical environments. 
$k_{col}^{-1}$, direct collisional rate from the 
ground level (Green et al. \cite{Green93}).
$k_{rad}^{-1}$, radiative excitation rate via the $v2\ 1_{10}$ vibration state 
(absorption of a $6.18 \mic$ photon). 
}

\begin{flushleft}
\begin{tabular}{lccccc}
\hline
  &  $<I_{\nu}>$ & T & $\nHdeux$& $k_{col}^{-1}$ & $k_{rad}^{-1}$\\
  & $MJy/sr$ & K & $\cmcube$ & $\smun$ & $\smun$ \\
  \hline
 PDR & 3 $\dix^4$ & 120 & $\dix^{4}$ & 6.4~$\dix^{-9}$ & 1.8~$\dix^{-6}$ \\
M17 SW & $1.5 \dix^4$ & 50 & $\dix^5$ & 3.7~$\dix^{-9}$  &0.9~$\dix^{-6}$\\
 Starburst & $\dix^4$ & 120 & $\dix^4$ & 6.4~$\dix^{-9}$& 6.0~$\dix^{-7}$\\
Gal. $\Htwo$& 40 & 50 & $\dix^4$ & 3.7~$\dix^{-10}$& 2.4~$\dix^{-9}$ \\
 \hline
 \end{tabular}
 \end{flushleft}
 \end{table}

\section{\label{Model} The model}

Our model solves the equations of statistical equilibrium 
for a given molecule, including rotation
levels in the ground and first excited vibration states, 
assuming an homogenous gas cloud
(ie the excitation of the molecule doesn't vary across the cloud) :

\begin{eqnarray}
\frac{dx_i}{dt} &=&  -x_i \sum_{j \ne i}(C_{ij} \nHdeux + R_{ij})
+ \sum_{j \ne i} x_j (C_{ji} \nHdeux + R_{ji})\nonumber \\
&=& 0
\label{eq:statequ}
\end{eqnarray}

where $x_i$ is the molecular fraction in each energy level ($\sum_{i}x_i = 1$),
$\mbox{C}_{ij}$ ($\cmcubes$) are the collision rates, 
$\mbox{R}_{ij}$ ($\smun$) the radiative rates and $\nu_{ij}$ the photon
frequency of the 
transition. The radiative transition rates read:

\begin{eqnarray}
R_{ij} & = & A_{ij} (1+\frac{c^2}{2h\nu_{ij}^3}<I_{\nu}>)   \quad  for \ i > j
\nonumber \\
R_{ij} & = & A_{ji} \frac{c^2}{2h\nu_{ji}^3} \frac{g_j}{g_i} <I_{\nu}>  \quad  for \ i < j
\label{eq:radrates}
\end{eqnarray}

$<I_{\nu}>$ is the averaged spectral density of the 
radiation field at the frequency of the transition.
This average has to be performed both over the velocity
distribution of the molecules, and over the solid angles. 

Concerning the radiative transfer, we assume that the PDR is a clumpy medium at small 
scale, with molecular clumps embedded in a lower density interclump gas, 
$n_H = 10^3$ to 
$10^4 \cmcube$. There are several observational evidences which support this hypothesis, 
among which one can cite: 1/ the high CI and CII column densities measured toward the 
M17-SW PDR and molecular cloud by Genzel et al. \cite{Genzel88} and Stutzki et al. 
\cite{Stutzki88}, and 2/ the extension of the PAH emission in the Orion bar, which 
shows that the UV flux penetrates much deeper into the molecular cloud than what is 
allowed by extinction in a homogeneous gas (see Tielens et al.\cite{Tielens93} and Giard 
et al. \cite{Giard94b}). We assume in the following that the UV radiation is converted
to near infrared PAH photons in the 
interclump medium and the outer parts of the molecular 
clumps, the molecules being at larger Av behind the dissociation limit.
The peculiarity of our problem, compared 
to the usual line transfer within a dust cloud (see Takahashi et al. \cite{Takahashi83} and 
Collison and Watson \cite{Collison95}), is that the dust component responsible for the 
near infrared exciting radiation (the PAHs) is not at thermal equilibrium. Here we didn't 
try to take into account the absorption and emission by dust  in a selfconsistent way. 
Instead, we have neglected the absorption by dust and we use for the dust radiation field 
the model of D\'esert et al. \cite{Desert90}, with the flux scaled on the 3.3 $\mic$ 
measurements of Giard et al. \cite{Giard94b} on M17 South-West and the Orion bar.
The only extinction is then the one of the molecules into the clumps.
We can use for the dust photons a radiative transfer of the form $(1-exp(-\tau))/\tau$, which has
been shown to be a good approximation for a medium with absorbing clumps and transparent
intercloud phase (see Appendix A in Giard et al. 
\cite{Giard92}). However, the uncertainty introduced in the model
by the transfer used for the dust photons is attenuated by the 
integration over the line profile which follows in
Equ. \ref{eq:beta} below.

With such hypothesis the transfer equation reads:

\begin{eqnarray}
I_\nu = S (1-exp(-\tau)) + F_d \frac{(1-exp(-\tau))}{\tau}  
\label{eq:radtra}
\end{eqnarray}

where $F_d$ is the dust and PAHs spectral density in the continuum, 
and S is the line source function:

\begin{equation}
S = \frac{2h\nu^3}{c^2} \frac{1}{\frac{g_u x_l}{g_l x_u}-1}
\label{eq:source}
\end{equation}

u and l denote the upper and lower level of the transition, 
and $\tau$ is the wavelength dependent molecular optical depth:

\begin{equation}
\tau(\nu) = N x_u \phi(\nu) \frac{\Aul}{4\pi} \frac{h\nu}{S}
\label{eq:tau}
\end{equation}

N is the molecular column density, and $\phi(\nu)$ the normalized doppler
profile resulting from the velocity distribution of the molecules. 
In the most general case of our aproximations, the profile averaged radiation
density
used in Equ. \ref{eq:statequ} (statistical equilibrium) reads:

\begin{equation}
<I_\nu> \simeq S (1 - \beta_m) + \frac{F_d}{2} \beta_d
\label{eq:Jnu}
\end{equation}

where the escape probabilities for molecule and dust photons read:

\begin{eqnarray}
\beta_m(<\tau>) & =  & \int exp(-\tau(\nu)/2) \phi(\nu) d\nu \nonumber \\
\beta_d(<\tau>) & = & \int \frac{1 - exp(-\tau(\nu)/2)}{\tau(\nu)/2} \phi(\nu) d\nu  
\label{eq:beta}
\end{eqnarray}

The factor 2 in Equ. \ref{eq:Jnu} and \ref{eq:beta} arises from the fact that, 
on average,
the column density surrounding any point within the cloud is half that
seen by the observer.

\begin{figure}[t]
\vbox to 5.5cm{
\epsfbox{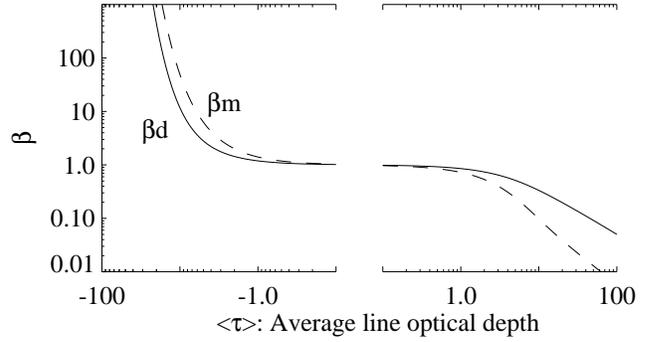}
}
\caption[]{Escape probabilities for molecule and dust photons versus 
the line average optical depth.
\label{figbeta}}
\end{figure}

$\beta_m$ and $\beta_d$ have been computed numerically for a Maxellian 
velocity distribution. They are plotted in Fig. \ref{figbeta}
as a function of the average line optical depth:

\begin{equation}
<\tau> = \frac{N}{DV} \frac{c}{\nu} x_u \frac{\Aul}{4\pi}\frac{h\nu}{S} 
\label{eq:avgtau}
\end{equation}

where $DV$ is the width of the projected velocity distribution and S
the source function.

Finally, the integrated intensity of the lines are computed by evaluating
the power radiated by the molecules. Normalized to the velocity width it
reads:

\begin{eqnarray}
\frac{P_{ul}}{Wm^{-2}sr^{-1}(km/s)^{-1}} 
= \frac{N}{DV}\frac{h\nu}{4\pi} (x_u R_{ul} - x_l R_{lu}) \nonumber \\
= \frac{\nu}{c} <\tau> (\beta_m S - \frac{F_d}{2}\beta_d)
\label{eq:Inu}
\end{eqnarray}

\begin{table}[t]
\caption[]{\label{table:h2olevels}
Calculated population fractions  of the lower rotational levels of $\HtwoO$
(T = 120K, $\nHdeux = 10^4 \cmcube$, $N_{\HtwoO} = 10^{16} \cmdeuxkms$, 
ortho/para = 3). 
Column 4: ratio of the level population fraction to the population obtained
with a pure collisional model. 
}
{\bf  a) $\orthoHtwoO$}
\begin{flushleft}
\begin{tabular}{rllc}
\hline
  $Energy$ &    state   &   fraction     &fraction\\
  $[\cmun]$  &$~v~~$ $J K^{-}K^{+}$    &in state   & /fraction(Coll.)  \\
\hline
       23.8&000  1  0 1&  8.75e-01&       1.0\\
       42.4&000  1  1 0&  9.18e-02&       0.9\\
       79.5&000  2  1 2&  3.21e-02&       2.5\\
      134.9&000  2  2 1&  2.59e-04&     100.3\\
      136.8&000  3  0 3&  3.58e-04&      25.9\\
      173.4&000  3  1 2&  1.53e-05&      66.4\\
      212.2&000  3  2 1&  1.20e-05&     225.5\\
      224.8&000  4  1 4&  1.19e-07&       4.8\\
      285.4&000  3  3 0&  4.47e-08&     656.7\\
\hline
\end{tabular}
\end{flushleft}
{\bf  b) $\paraHtwoO$}
\begin{flushleft}
\begin{tabular}{rllc}
\hline
   $Energy$ &    state   &   fraction     &fraction\\
  $[\cmun]$  &$~v~~$ $J K^{-}K^{+}$    &in state   & /fraction(coll.)  \\
\hline
       0.0&000  0  0 0&  8.68E-01&       0.9\\
       37.1&000  1  1 1&  1.19e-01&       2.0\\
       70.1&000  2  0 2&  1.18e-02&      15.7\\
       95.2&000  2  1 1&  6.35e-04&     237.6\\
      136.2&000  2  2 0&  5.63e-05&     136.9\\
      142.3&000  3  1 3&  1.08e-05&      37.7\\
      206.3&000  3  2 2&  1.35e-07&     251.4\\
      222.1&000  4  0 4&  2.55e-08&       1.2\\
      275.5&000  4  1 3&  1.12e-09&       2.7\\
      285.2&000  3  3 1&  2.63e-08&     648.0\\
\hline
\end{tabular}
\end{flushleft}
\end{table}

We have run the model for $\HtwoO$ and $\NHtrois$, 
two polar molecules likely to be abundant 
in PDRs because they are expected to be formed in grain mantles and released
to the gas in such regions (see e.g. Shalabiea and Greenberg \cite{Shalabiea94}). 
The physical parameters of the regions are typical of a PDR
in a region of high mass star formation (first line in Table \ref{firstestimates}).
Concerning the dust spectrum, we have used the 
model of D\'esert et al. (\cite{Desert90}) in the case of dust irradiation by  
an O star. The dust flux has been scaled so that the peak 6.2 $\mic$
brightness
is $3~10^4$ MJy/sr. This is the value expected for this type of PDRs,
on the basis of the 3.3 $\mic$ feature flux measured toward M17 south-west
and the Orion bar (Giard et al. \cite{Giard94b}).The ISO-SWS spectrum
of Verstraete et al (\cite{Verstraete96}) on the interface region in M17
is about a factor of 15 below the above surface brightness, because of
dilution in the  ISO-SWS aperture ($14\arcsec \times 20 \arcsec$) and averaging  
over 3 different positions.

The coupled system of Equs. \ref{eq:statequ} and \ref{eq:Jnu}, statistical
equilibrium and radiative transfer, have been solved numerically for each
molecule, varying the hydrogen density and the molecule column density.
We selected the 16 lower rotation levels in each vibration state
of the molecule studied. The collision deexcitation rates for transitions 
between rotation levels in the ground vibration state are taken from
Green et al. (\cite{Green93}) for $\HtwoO$, and Danby et al. (\cite{Danby88})
for
$\NHtrois$. Concerning the collision deexcitation rates between different vibration
states, we have assumed the rate value of the 
same rotation states within the fundamental vibration state. Collision
rates values are not critical since, for the temperature and
densities considered, T $=$ 120K, $\dix^2\cmcube < \nHdeux < \dix^7 \cmcube$, 
the transitions between two different
vibration states are largely dominated by the radiative rates.
The line frequencies and the Einstein coefficients are taken from the GEISA data
bank 
(see Husson et al. \cite{Husson92}). As each of the two molecules considered,
$\HtwoO$ and
$\NHtrois$, have an ortho and para configuration which are not
coupled, we have actually run the model for four different molecular species.
We have considered two vibration states for $\HtwoO$:  000 (0 $\cmun$)
  and 010 (1620 $\cmun$),  
three in the case of $\NHtrois$: 0000 (0 $\cmun$), 0100 (950 $\cmun$)
and 0001 (1637 $\cmun$).

\begin{table}[t]
\caption[]{\label{table:nh3levels}
Calculated population fractions  of the lower rotational levels of $\NHtrois$
(T = 120K, $\nHdeux = 10^4 \cmcube$, $N_{\NHtrois} = 10^{15} \cmdeuxkms$, 
ortho/para = 2). Same as Table \ref{table:h2olevels}. 
}
{\bf  a) $\orthoNHtrois$}
\begin{flushleft}
\begin{tabular}{rllc}
\hline
  $Energy$ &    state   &   fraction    &fraction\\
  $[\cmun]$  &$~~v~~$ $J K\pm$    &in state   & /fraction(coll.)  \\
\hline
       0.4&0000  0 0-&  2.59e-01&       0.9\\
       19.5&0000  1 0+&  7.95e-02&       1.8\\
       60.0&0000  2 0-&  1.19e-03&      10.6\\
       85.9&0000  3 3-&  3.04e-01&       1.0\\
       86.7&0000  3 3+&  3.07e-01&       1.0\\
      118.8&0000  3 0+&  1.57e-06&       3.8\\
      165.3&0000  4 3-&  4.09e-04&      11.2\\
      166.1&0000  4 3+&  3.89e-04&      18.0\\
\hline
\end{tabular}
\end{flushleft}
{\bf  b) $\paraNHtrois$}
\begin{flushleft}
\begin{tabular}{rllc}
\hline
  $Energy$ &    state   &  fraction  &fraction\\
  $[\cmun]$  &$~~v~~$ $J K\pm$    &in state   & /fraction(coll.)  \\
\hline
       16.2&0000  1 1-&  2.65e-01&       1.0\\
       17.0&0000  1 1+&  2.45e-01&       1.0\\
       44.8&0000  2 2+&  1.97e-01&       1.0\\
       45.6&0000  2 2-&  1.80e-01&       1.0\\
       55.9&0000  2 1-&  9.80e-04&       3.2\\
       56.7&0000  2 1+&  1.01e-03&       3.2\\
      104.4&0000  3 2+&  1.60e-04&      10.1\\
      105.2&0000  3 2-&  1.81e-04&      10.4\\
      115.5&0000  3 1-&  2.88e-06&       3.9\\
      116.3&0000  3 1+&  2.61e-06&       3.9\\
\hline
\end{tabular}
\end{flushleft}
\end{table}

\section{\label{Results} Results}

The population fraction of the rotation levels in the
ground vibration state are presented in Tables \ref{table:h2olevels} a, b
and \ref{table:nh3levels} a, b
for column densities of $10^{16} \cmdeuxkms$ ($\HtwoO$)
and  $10^{15} \cmdeuxkms$ ($\NHtrois$), and statistical ortho to para
ratio of 3:1 for $\HtwoO$ and 2:1 for $\NHtrois$. 
Given the typical hydrogen column density
of a PDR, $2N_{\Htwo} \simeq 10^{21} \cmdeuxkms$
($2 < A_V < 5$ over 5 km/s, see Tielens and Hollenbach \cite{Tielens85}), 
such molecular column densities correspond to abundances of 
$10^{-5}$ and $10^{-6}$ for $\HtwoO$ and $\NHtrois$ respectively.

\clearpage

\begin{table*}[t]
\caption[]{\label{table:h2olines}
Calculated intensities  of the brighter $\HtwoO$ lines.
(T = 120K, $\nHdeux = 10^4 \cmcube$, $N_{\HtwoO} = 10^{16} \cmdeuxkms$, 
ortho/para = 3). 
Column 1: wavelength in micrometer; column 2: Einstein coefficient; 
column 3: line intensity; column 4: ratio of the level population fraction to 
the population obtained  with a pure collisional model.
}
\begin{flushleft}
\begin{tabular}{rlccll}
\hline
  $\lambda$ &$A_{ij}$& $F$ &$F$& Upper & Lower\\
  $[\mic]$  &$[s^{-1}]$&$[Wm^{-2}sr^{-1}(km/s)^{-1}]$& /F(coll.)
  &$~v~~$ $J K^{-} K^{+}$&$~v~~$ $J K^{-} K^{+}$ \\
\hline
\\
\multicolumn{6}{l}{{\bf  a) $\orthoHtwoO$}}\\
   538.289& 3.6e-03&  2.51e-10&    0.9& 000  1  1 0&000  1  0 1\\
   273.193& 1.7e-02&  1.00e-09&   59.2& 000  3  1 2&000  3  0 3\\
   259.984& 3.9e-03&  2.63e-10&   64.7& 000  3  1 2&000  2  2 1\\
   257.791& 2.4e-02&  1.30e-09&  225.2& 000  3  2 1&000  3  1 2\\
   180.488& 3.2e-02&  2.21e-09&   25.9& 000  2  2 1&000  2  1 2\\
   179.526& 9.7e-02&  3.37e-09&    2.3& 000  2  1 2&000  1  0 1\\
   174.626& 7.4e-02&  1.89e-09&    7.3& 000  3  0 3&000  2  1 2\\
    75.380& 4.8e-01& -4.17e-09& -119.2& 000  3  2 1&000  2  1 2\\
     6.642& 7.4e+00&  4.51e-08&       & 010  1  1 0&000  2  2 1\\
     6.185& 1.0e+01& -4.84e-08&       & 010  1  1 0&000  1  0 1\\
\hline
\\
\multicolumn{6}{l}{{\bf  b) $\paraHtwoO$}}\\
   398.643& 7.3e-03&  1.07e-09&   62.1& 000  2  1 1&000  2  0 2\\
   303.459& 1.0e-02&  1.87e-09&    7.4& 000  2  0 2&000  1  1 1\\
   269.273& 5.7e-02&  1.50e-09&    2.1& 000  1  1 1&000  0  0 0\\
   243.972& 1.9e-02&  1.68e-09&  129.5& 000  2  2 0&000  2  1 1\\
   138.527& 1.8e-01&  1.98e-10&    1.6& 000  3  1 3&000  2  0 2\\
   100.983& 4.5e-01& -2.32e-09& -226.0& 000  2  2 0&000  1  1 1\\
    67.089& 1.8e+00&  1.81e-10&  423.5& 000  3  3 1&000  2  2 0\\
    46.484& 3.0e-02& -2.47e-10& -2.4e+04& 000  3  3 1&000  2  0 2\\
     6.672& 6.3e+00&  2.34e-08&        & 010  1  1 1&000  2  2 0\\
     6.390& 3.0e+00&  1.16e-08&        & 010  1  1 1&000  2  0 2\\
     6.116& 2.1e+01& -3.80e-08&       & 010  1  1 1&000  0  0 0\\
\hline
\end{tabular}
\end{flushleft}
\end{table*}

\begin{table*}[h]
\caption[]{\label{table:nh3lines}
Calculated intensities  of the brighter $\NHtrois$ lines.
(T = 120K, $\nHdeux = 10^4 \cmcube$, $N_{\NHtrois} = 10^{15} \cmdeuxkms$, 
ortho/para = 2). Same as Table \ref{table:h2olines}.
}
\begin{flushleft}
\begin{tabular}{rlccll}
\hline
  $\lambda$ &$A_{ij}$& $F$ &$F$& Upper & Lower\\
  $[\mic]$  &$[s^{-1}]$&$[Wm^{-2}sr^{-1}(km/s)^{-1}]$& /F(coll.)
  &$~~v~~$ $J K\pm$&$~~v~~$ $J K\pm$ \\
\hline
\\
\multicolumn{6}{l}{{\bf  a) $\orthoNHtrois$}}\\
   523.670& 9.5e-03&  3.15e-10&    2.2& 0000  1 0+&0000  0 0-\\
   246.773& 6.1e-02&  6.09e-10&    4.6& 0000  2 0-&0000  1 0+\\
   169.996& 1.7e-01&  1.17e-10&    2.8& 0000  3 0+&0000  2 0-\\
   127.108& 8.4e-02&  2.92e-10&    2.6& 0000  4 3-&0000  3 3+\\
   124.648& 8.8e-02&  2.71e-10&    3.7& 0000  4 3+&0000  3 3-\\
    11.209& 1.0e+01&  9.09e-09&       & 0100  1 0+&0000  2 0-\\
    10.507& 3.1e+01& -9.64e-09&       & 0100  1 0+&0000  0 0-\\
\hline
\\
\multicolumn{6}{l}{{\bf  b) $\paraNHtrois$}}\\
   256.575& 2.0e-02&  1.20e-10&    2.2& 0000  2 1-&0000  1 1+\\
   246.694& 2.3e-02&  1.25e-10&    2.2& 0000  2 1+&0000  1 1-\\
   169.990& 7.5e-02&  5.92e-11&    3.5& 0000  3 1-&0000  2 1+\\
   169.967& 4.7e-02&  8.91e-11&    2.7& 0000  3 2+&0000  2 2-\\
   165.728& 8.1e-02&  5.92e-11&    3.5& 0000  3 1+&0000  2 1-\\
   165.596& 5.0e-02&  9.08e-11&    2.6& 0000  3 2-&0000  2 2+\\
    11.460& 4.5e+00&  6.04e-10&       & 0100  2 1-&0000  3 1+\\
    11.011& 5.1e+00&  6.46e-10&       & 0100  2 1+&0000  3 1-\\
    10.289& 8.2e+00& -1.03e-09&       & 0100  2 1-&0000  1 1+\\
     9.925& 9.2e+00& -1.13e-09&       & 0100  2 1+&0000  1 1-\\
\hline
\end{tabular}
\end{flushleft}
\end{table*}

\clearpage

Except the peculiar case of the $\NHtrois$ 
metastable levels (J = K, see Sect. \ref{nh3}), the population fractions are far
from
thermal equilibrium, because the density used does not allow
collisional excitation rates to overwhelme spontaneous radiative emission
(critical $\Htwo$ density $\simeq 10^8 \cmcube$ for excitation
of a 100 $\cmun$ level of $\NHtrois$ or $\HtwoO$ in a 100K gas). 
For this reason we compare the results of our model to
 a pure collisional model
where the dust flux has been set to zero (Column 4 in Tables 
\ref{table:h2olevels}a, b
 and \ref{table:nh3levels}a, b).

The predicted intensities of the brightest $\HtwoO$ and $\NHtrois$ lines are 
listed in Tables \ref{table:h2olines}a,b and \ref{table:nh3lines}a,b. 
Fig. \ref{fig:totalspectrum} and \ref{fig:totalspectrum_nh3}
display the calculated synthetic infrared spectra emerging from 
the PDR. 
In these plots, the height of each 
line above or below the continuum is not proportional to its integrated intensity,
because the different optical depths of the lines induce different
line widths.
The insert shows the details of the expected NIR vibration transitions. 

Our results show that the net effect
of the NIR pumping is to slightly reduce the fraction of molecules 
in the lowest rotation level, increasing the population of the excited levels
and thus increasing the intensities
of the submillimeter and far infrared lines. 
However, the detailed behaviours for 
$\HtwoO$ and $\NHtrois$ are not the same, because of the different radiative 
selection rules.

\subsection{\label{h2o} $\HtwoO$}

\begin{figure}[t]
\vbox to 6cm{
\epsfbox{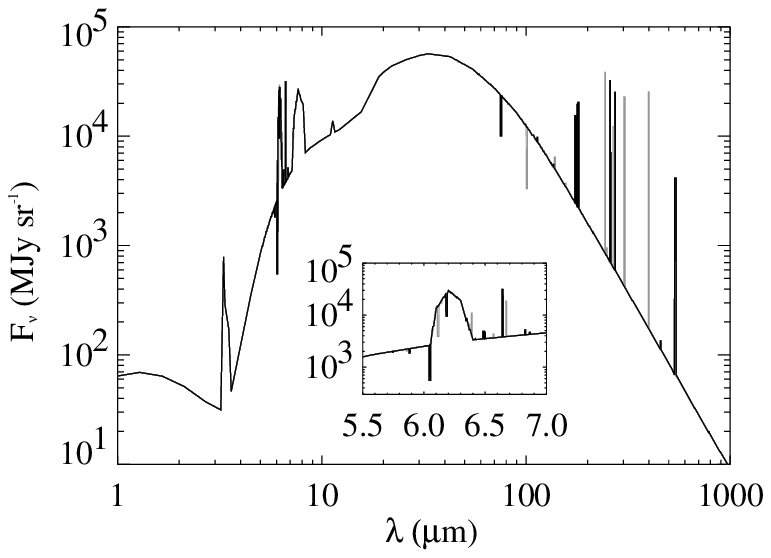}
}
\caption[]{Synthetic dust plus $\HtwoO$ spectrum emerging
from the PDR (dashed lines for para species). 
T = 120K, $\nHdeux = 10^4 \cmcube$,
$N_{\HtwoO} = 10^{16} \cmdeuxkms$, ortho/para = 3.
\label{fig:totalspectrum}}
\end{figure}

\begin{figure}[t]
\vbox to 15cm{
\epsfbox{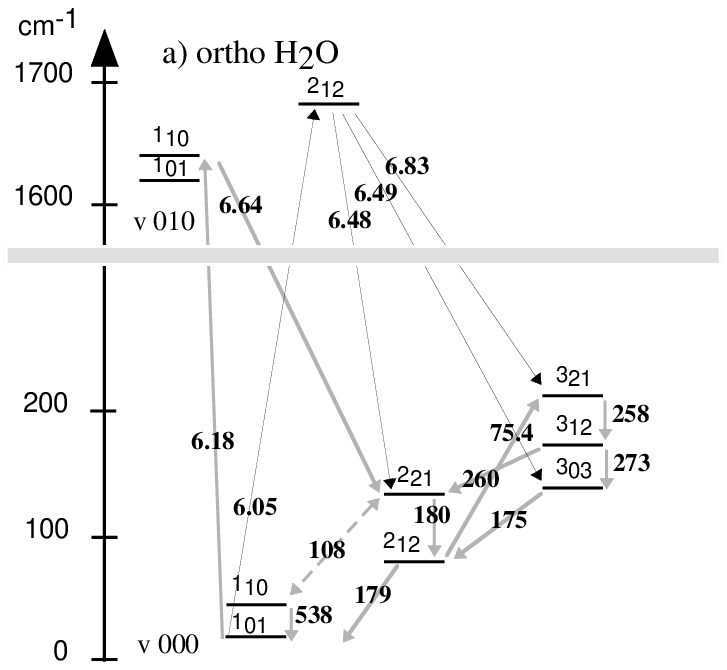}
\epsfbox{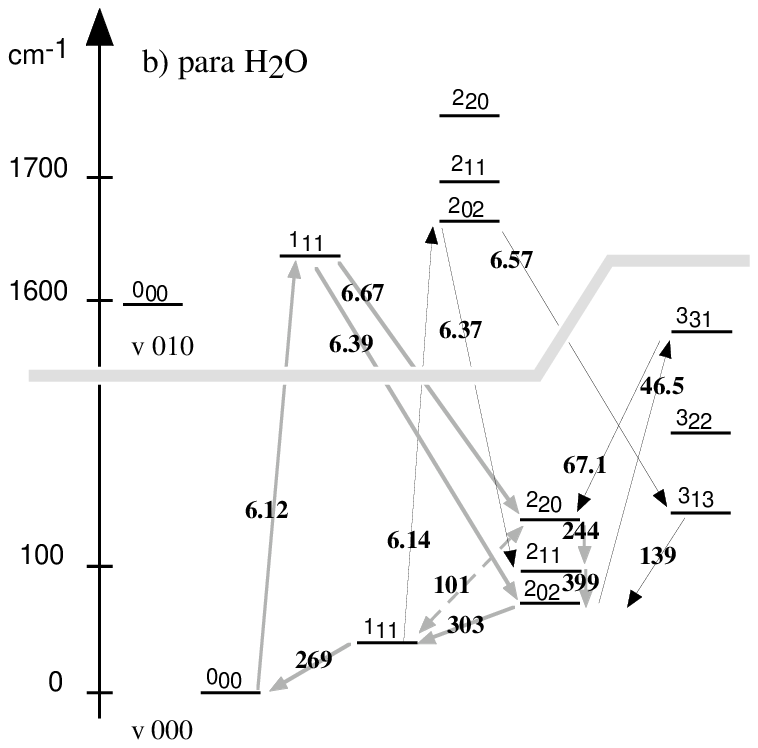}
}
\caption[]{Levels and main radiative transitions for a) ortho and b) para
$\HtwoO$.
The wavelength of each transition in micron is indicated in bold characters.
(Thick arrows for the stronger transitions).
\label{fig:h2olevels}}
\end{figure}

$\HtwoO$ is an asymmetric top
with a dipole moment of 1.94 Debye. 
We refer to the rotation levels of $\HtwoO$ using the general $J_{K^- K^+}$
numbering 
of the asymmetric top molecules, where $J$ is the total angular momentum 
and $K^-$ (respectively  $K^+$)  is the projection of the angular 
momentum on the symmetry axis for the limiting prolate (respectively oblate)
molecule 
(see Townes and Shallow \cite{Townes75}).
The selection rules are, 
$\Delta J = 0, +1, -1$ and $\Delta( K^{-}-K^{+})$ even. The later
rule forbids para/ortho radiative transitions since $K^{-}-K^{+}$ is odd for
ortho and even for $\paraHtwoO$. 

As can be seen from Fig. \ref{fig:h2olevels}a, 
the excitation scheme for $\orthoHtwoO$ is quite complex
because of the combination of the radiative pumping by NIR photons in the
vibrational transitions and FIR photons in the rotational transitions.
The molecule is mostly excited by NIR
pumping in the $v010$ $1_{10}$ level and subsequent deexcitation in
the $2_{21}$ rotation level of the vibration ground state 
(absorption at 6.18 $\mic$ and emission 
at 6.64 $\mic$). The population fraction of the $v000$ $2_{21}$ level
is considerably increased with respect to the comparison model ($\simeq 100$). 
Two different de-excitation channels are possible from this level. The
first one through the 108 $\mic (2_{21}-1_{10})$ line, and the
second one through the 180.5 $\mic (2_{21}-2_{12})$ line. The first channel 
is not efficient because for the parameters
we have selected, $\nHtwo = 10^4 \cmcube$ and $\frac{N_{\HtwoO}}{DV} =
10^{16} \cmdeuxkms$, emission in the 108 $\mic$ line
is balanced by radiative pumping by FIR dust photons in the 
same line. Thus, the de-excitation occurs mainly 
trough the 180.5 $\mic$ line. The intensity of this line is increased
by a factor of order 25 with respect to the collision model. Once
in level $2_{12}$, the molecule may either be excited in the $3_{21}$ level by
absorption of a 75.4 $\mic$ dust photon, or de-excited in the ground level
($1_{01}$) by emission at 179.5 $\mic$. The population fractions of 
the J=3 ladder are increased by one to two decades with respect to the collisional 
model. Consequently, the intensities of the FIR and submillimeter de-excitation 
lines are enhanced by similar factors. NIR pumping at 6.18 $\mic$ has 
thus a major effect on the $(2_{21}-2_{12}$)
line at 180.5 $\mic$, whereas FIR pumping at 75.4 $\mic$ mostly enhances the  
following transitions: $(3_{21}-3_{12}$)
at 258 $\mic$, $(3_{12}-2_{21}$) at 260 $\mic$, 
$(3_{12}-3_{03}$) at 273 $\mic$ , and $(3_{03}-2_{12}$) at 175 $\mic$.

\begin{figure*}[t]
\vbox to 6.5cm{
\epsfbox{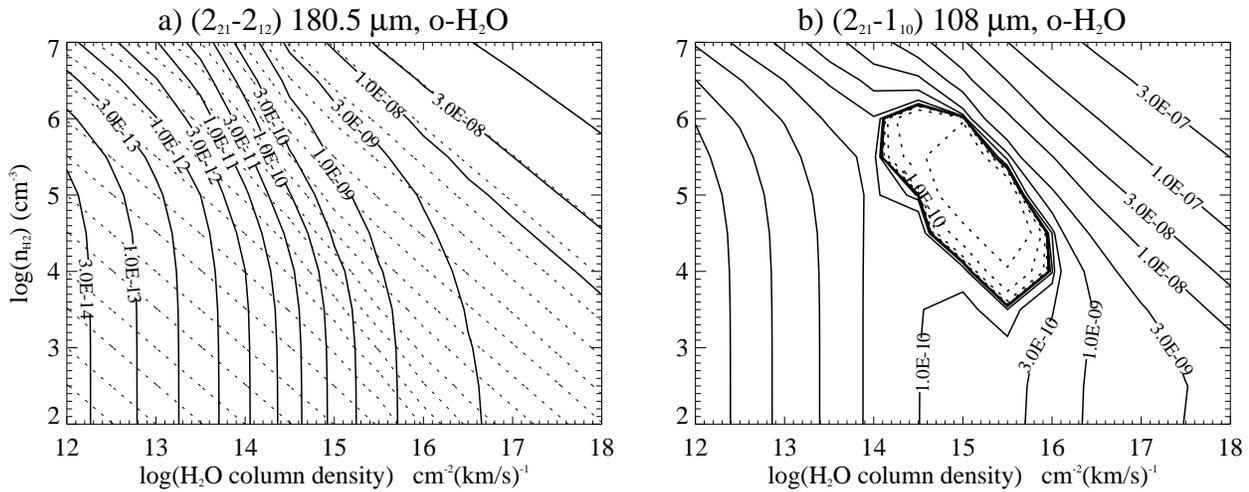}
}
\caption[]{ iso-intensities of a) $\orthoHtwoO$ ($2_{21}-2_{12}$) 180.5 $\mic$ 
and b) $\orthoHtwoO$ ($2_{21}-1_{10}$) 108 $\mic$ lines 
as a function of the molecule column density and the hydrogen
density. Dotted lines in a) are the contours obtained with a pure 
collisional model.
Dotted lines in b) for an absorption line. Units are $Wm^{-2}sr^{-1}(km/s)^{-1}$.
 \label{fig:coldensdiag}}
\end{figure*}

The intensity of the 108 $\mic$ line is driven both by NIR and FIR pumping,
and this line can appear in emission or absorption against the dust continuum,
depending on the radiation field, the gas density and the molecular column
density (see Fig\ref{fig:coldensdiag}b). 

The case of $\paraHtwoO$ is quite similar to $\orthoHtwoO$. The main
NIR absorption 
line, ($0_{00}-1_{11}$),is located at 6.12  $\mic$, with re-emissions predicted at 
6.39 and 6.67 $\mic$ (see Fig. \ref{fig:h2olevels}b), populating mainly the J=2
rotation ladder. Excitation of the J=3 ladder is much less efficient than in 
the case
the ortho species because the allowed transition has a low Einstein 
coefficient ($3~10^{-2}$ against  0.48 for the 75.4 $\mic$ line of 
$\orthoHtwoO$). NIR and FIR pumping are thus responsible for bright 
de-excitation lines in the J=2 ladder: $(2_{20}-2_{11})$ at 244 $\mic$,
$(2_{11}-2_{02})$ at 399 $\mic$ and $(2_{02}-1_{11})$ at 303 $\mic$.
As for the $\orthoHtwoO$ 108 $\mic$ line, the $(2_{20}-1_{11})$ line
at 101 $\mic$ will
appear either in absorption or emission on the continuum, depending on
the physical parameters of the region.

In the regime where the NIR pumping line is not saturated ( 
$N_{H2O} < 10^{16} \cmdeuxkms$) the absorptions at 6.18 and 6.12 $\mic$ 
give a direct estimate of the ortho and para column densities.
Furthermore, FIR lines will increase with the pumping rate, i.e. the
radiation field. This later parameter can be fairly 
well known by direct observation
of the
dust and PAH NIR spectrum. Once this is done, the intensities of the FIR lines
can be used to trace 
the molecule column densities. This is shown in Fig. \ref{fig:coldensdiag}a
where the calculated iso-intensities of the 180.5 $\mic$  ($2_{21}-2_{12}$) line
are drawn versus the $\HtwoO$ 
column density and the $\Htwo$ density. The dotted contours in the same plot 
correspond to the 
comparison model (collisions only). In the comparison model the line
intensity increases fairly regularly with the gas density and the molecule column
density. If radiative pumping is introduced as in our model, it dominates the
excitation at moderate densities, $\nHtwo < 10^5 \cmcube$, so long as the 6.18
$\mic$ pumping line is optically thin: $N_{\HtwoO} < 10^{16} \cmdeuxkms$. 
In this domain of the
parameter space, which is appropriate for the conditions that prevail within PDRs, 
the line intensity is almost independant of the gas density and can be
used to trace the column density.  
However, for the 180.5 $\mic$ line, the intensity is not a linear function of the 
column density. This is 
because the upper and lower levels of the transition have several other radiative
exciting or deexciting channels which become optically thick in different 
column density ranges.

Fig. \ref{fig:coldensdiag}b shows the intensity contours of the   
($2_{21}-1_{10}$) 108 $\mic$ $\orthoHtwoO$ line. For the sake of
clarity we didn't overplot the contours obtained with the comparison model. They
are quite similar to those of the 180.5 $\mic$ line in
 Fig. \ref{fig:coldensdiag}a. 
As mentionned above, this line appears either
in absorption or emission, depending on the gas density, the molecule column
density and the radiation field. Including NIR pumping favors emission in 
this line since the upper level is directly fed by the de-excitation of the 
excited vibration state. This is why the domain
where the line will appear in absorption against the dust continuum
is restricted to a small region: $10^{14} < N_{\HtwoO} < 10^{16} \cmdeuxkms$, 
$10^4 < \NHdeux < 10^6 \cmcube$. 

It is worth pointing out that, in addition to increasing considerably
the $\HtwoO$ submillimeter and FIR line intensities, radiative pumping
will strongly modify the lines intensity ratios with respect to pure 
collisional models. For instance, the 180.5 $\mic$ to 179.5 $\mic$ $\orthoHtwoO$
line intensity ratio is increased by a factor of order 10, and
the 398 $\mic$ $\paraHtwoO$ to 538 $\mic$ $\orthoHtwoO$ submillimeter
lines ratio is enhanced by a factor of 60. Such effects have very important
observational consequences, particularly when the lines are used to derive
column densities and ortho to para abundance ratios.

Finally, the NIR lines predicted to appear in absorption and emission 
against the PAH continuum can directly be used to probe the 
$\HtwoO$ column density. The line to continuum contrast of the emission
transitions is high (see insert in Fig. \ref{fig:totalspectrum}), 
because the 
intensity of these lines comes from energy absorbed at the maximum 
of the 6.2 $\mic$
PAH band. With $\orthoHtwoO$, we actually have a case of quasi-resonant
excitation. The primary emission transition at 6.64 $\mic$ 
is thus a good  target for observations. For instance, if we assume a 
line width of 10 km/s, 
the measured line to continuum contrast is still of order 50$\%$ in 
the 6.64 $\mic$
transition, after dilution by the ISO-SWS grating spectral resolution 
($R \simeq 1500$). 
This is to be compared
to the contrast in the absorption line at 6.18 $\mic$, $\simeq 5 \%$ 
for the same
resolution.

\subsection{\label{nh3} $\NHtrois$}

\begin{figure}[t]
\vbox to 6cm{
\epsfbox{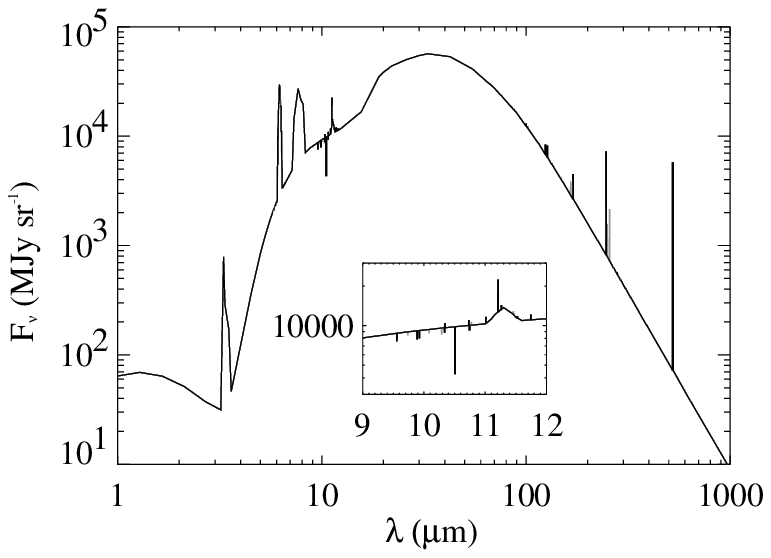}
}
\caption[]{ Synthetic dust plus $\NHtrois$ spectrum emerging
from the PDR. Same as Fig. \ref{fig:totalspectrum}. 
T = 120K, $\nHdeux = 10^4 \cmcube$,
$N_{\NHtrois} = 10^{15} \cmdeuxkms$, ortho/para = 2.
 \label{fig:totalspectrum_nh3}}
\end{figure}

\begin{figure}[t]
\vbox to 15cm{
\epsfbox{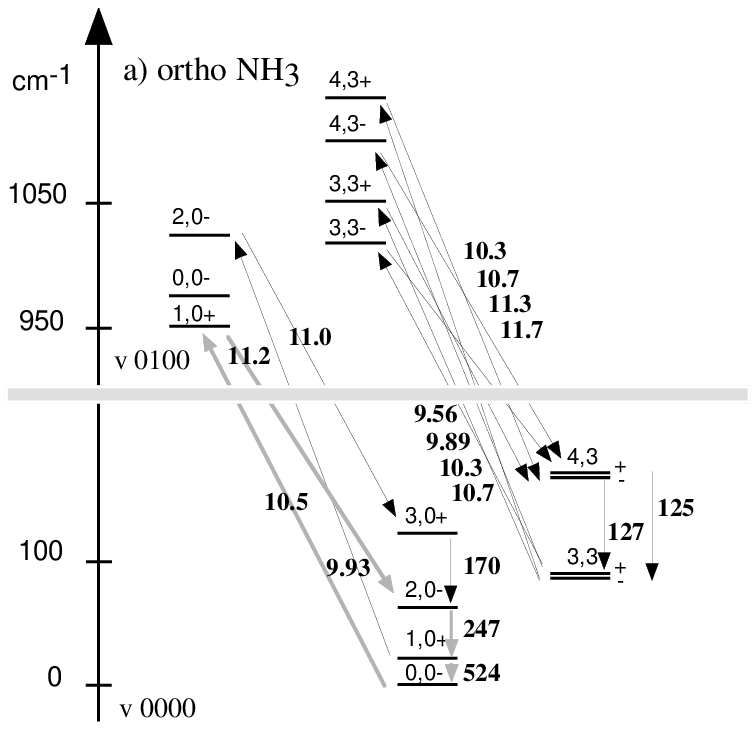}
\epsfbox{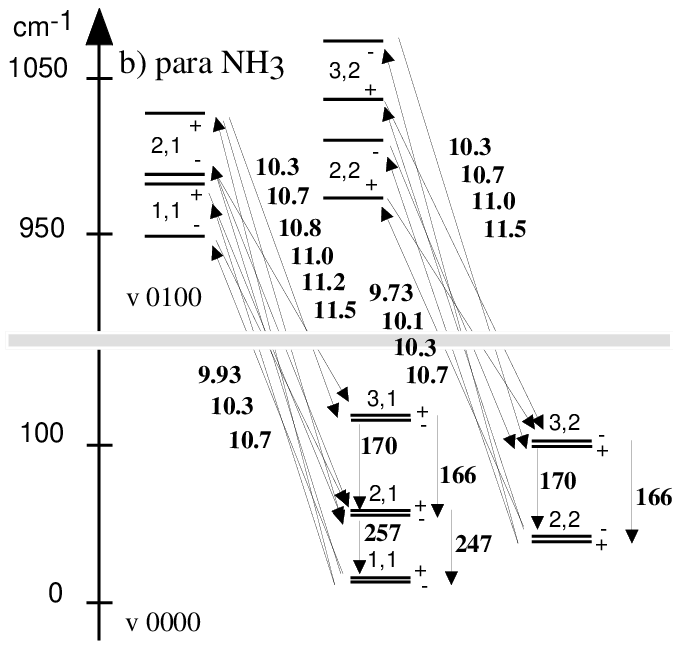}
}
\caption[]{Levels and main radiative transitions for ortho and para $\NHtrois$.
Same as Fig. \ref{fig:h2olevels}.
\label{fig:nh3levels}}
\end{figure}

$\NHtrois$ is a symmetric top prolate rotator with a dipole of  1.47 Debye.
The possible tunelling of 
the nitrogen nucleus through the hydrogen plan is responsible for the inversion 
splitting of the rotation levels with $K \ne 0$ (levels + and -).
We refer to the levels using the usual $J, K \pm$ numbering (see Townes 
and Shallow \cite{Townes75}). 
An additional quantum number 
due to a rotation-vibration interaction, 
$ l= \pm 1$, appears in the case of the degenerate vibration states (e.g. $0001$). 
The selection rules 
are those of a symmetric top: $\Delta J  = 0, +1, -1$;  $+\leftrightarrow-$; and $\Delta K =
0$ 
for transitions 
which do not involve a degenerate vibration state.  
The later rule implies that the different K ladders are not radiatively coupled.
Consequently, the lower + and - rotation levels of each K ladder (J = K) 
are the bottom 
of a radiative cascade,
and are almost in thermal equilibrium with the bottom 
levels of the other ladders. They are usually refered to as metastable
levels.

Three vibration states have been included in the excitation calculation
of $\NHtrois$:  0000, 0100, and 0001. However, despite comparable
Einstein coefficients, the frequency dependency of the radiative pumping
(see formula \ref{eq:k_rad}) favors radiative pumping in the $0100$ state 
($950 \cmun$)
with respect to the 0001 state ($1600 \cmun$).  
The $0100$ vibration at $950 \cmun$  correponds to the 
oscillation of the nitrogen nucleus above (or bellow) the hydrogens plan.

The NIR pumping induces a pump cycle within each K ladder, with 
several NIR absorption and emission transitions predicted between 9.5 and 12
$\mic$ (see Table \ref{table:nh3lines}). 
The intensities of the FIR and submillimeter deexcitation lines are 
increased by a factor ranging from 2 to 5: $(4,3 - 3,3)$ at 125 and
127 $\mic$, $(3,0 - 2,0)$ at 170 $\mic$, $(2,0 - 1,0)$ at 247 $\mic$
and $(1,0 - 0,0)$ at 524 $\mic$ for $\orthoNHtrois$; 
$(3,2 - 2,2)$,  $(3,1 - 2,1)$ at 166 and 170 $\mic$ and 
$(2,1 - 1,1)$ at 247 and 257 $\mic$ concerning $\paraNHtrois$. 

As for $\HtwoO$, the effect of NIR pumping is relatively  more efficient for the
transitions
from the higher rotation levels of the downward cascade, because
it is more difficult to populate these levels by collisions with the gas. 
However, in $\NHtrois$ the metastable levels are thermally populated
whatever is their energy. This facilitates the collisional excitation 
of high energy levels, if they are not far from the bottom of the K ladder .
For instance, levels (4,3)$\pm$ at 165 $\cmun$ are only 79 $\cmun$ above
the (3,3)$\pm$ levels from which they can be efficiently populated
by collisions. This explains why NIR pumping is relatively less efficient 
for $\NHtrois$ than $\HtwoO$.

Finally, there is no complex behaviour in the $\NHtrois$ lines as there
is in $\HtwoO$ (e.g. the 108 $\mic$ $\orthoHtwoO$ line). This is because
each K ladder is the location of a unique de-excitation cascade, forbidding
competition between different channels as in the $\HtwoO$ asymetric rotator. 

\section{\label{conclusion}Conclusion}

The main conclusions that can be drawn from our study are the following:

1) NIR photons are able to contribute very significantly to the rotational 
excitation of polar molecules within PDRs in galactic star-forming regions and 
starburst galaxies. 

2) As this mechanism is independent of the gas density,
we predict that bright FIR and submillimeter rotation lines should
be observed from the relatively moderate density and non shocked gas in PDRs, 
$\nHtwo < 10^5 \cmcube$,
where molecules are expected to be abundant because of grain mantles
evaporation.  

3) For $\HtwoO$ and $\NHtrois$ the intensity of the FIR and submillimeter 
rotation lines can be increased
by a factor ranging from 2 to several 100 with respect to models which do not
take into account NIR pumping. This means that the molecular column densities 
that 
can be infered from observations toward such regions, using standard models, 
can be wrong by similar factors. 

4) For gas densities below $10^5 \cmcube$, where NIR pumping dominates
over collisional excitation, the intensities of the FIR lines becomes mostly 
independant
of the gas density. This implies that the FIR and submillimeter 
lines can be used to probe
the molecule column density if the radiation field has been correctly estimated.

5) For $\orthoHtwoO$ we have a quasi-resonant excitation of the molecule,
by absorption of 6.2 $\mic$ PAH photons in the 6.18 $\mic$ fundamental
vibration transition. We thus predict 
a high line to continuum contrast for the emission de-excitation line
at 6.64 $\mic$.\\

NIR pumping is likely to affect not only the PDRs, as modelled in this paper,
but also the molecular clouds and cores in their vicinity. The extinction
in the 6 to 8 $\mic$ range is minimum (typically 0.02 of visible extinction,
Rieke and Lebofsky \cite{Rieke85}) and thus, the radiation produced in the
PDR penetrates well in the molecular clouds. 

Most molecules present in PDRs and nearby molecular
clouds will actually be affected by radiative NIR pumping. In addition
to $\HtwoO$ and $\NHtrois$ one can list
$CS$, $HCN$, $CO_2$, $CH_4$, $H_3O^+$, $HC_3N$, and $CH_3OH$. 
The combination of the absorption and emission lines from all species
existing in the gas phase is likely to be responsible 
for the very complex detailed structure of 
the 5 to 20 $\mic$ spectra measured at moderate spectral resolution 
in the direction of the M17-SW PDR with the ISO-SWS intrument
(Verstraete et al. \cite{Verstraete96}). Decrypting such
spectra in terms of molecular abundances will require the production
of synthetic model spectra including a significant number of molecules
of astrophysical interest. This will be done in a forthcoming paper.

\acknowledgements{
We are gratefull to B. Bonnet and N. Husson for making available to
us the full GEISA file.  
}

\end{document}